\documentclass[twoside]{article}
\usepackage{fleqn,espcrc2}

\usepackage{graphicx}
\usepackage[figuresright]{rotating}

\title{Superconductivity with the Meron-Cluster Algorithm}

\author{J.C. Osborn\address{Department of Physics, Box 90305, Duke University,
        Durham, NC 27708, U.S.A.}
	\thanks{This work was done in collaboration with S. Chandrasekharan
	under DOE grant DE-FG02-96ER40945.}}

\begin{document}

\begin{abstract}
The meron-cluster algorithm was previously used to extensively study
the physics associated with the spontaneous breaking of a discrete symmetry.
We recently discovered that a larger class of models with spontaneous breaking
of continuous symmetries can also be simulated using the meron-cluster
algorithm.
Here we study one of these new models which belongs to the attractive Hubbard
model family.
In particular we study the spontaneous breaking of the $U(1)$ fermion number
symmetry in two dimensions and find clear evidence for a
Kosterlitz-Thouless transition to a superconducting phase.
\end{abstract}

\maketitle

\section{Introduction}

For over a decade there has been a lot of interest in performing numerical
simulations of strongly correlated fermionic systems.
A recently discovered method for solving the sign problem inherent in such
simulations is the meron-cluster algorithm \cite{meron1}.
A new reference configuration allows us to explore a larger
class of models with this method.
The model we study here is a variant of the attractive Hubbard model.

The attractive Hubbard model has recieved a lot of attention as a toy model
for superconductivity.
It is a very simplistic model since the attraction between fermions is
explicitly included, instead of arising from some more subtle mechanism.
In that respect it is perhaps more closely related to the case of color
superconductivity in QCD.
It is still useful as a model for high-$T_c$ superconductors since at
intermediate attraction strength they both exhibit pairing at temperatures
above the superconducting transition \cite{Rand97}.

Despite the Hubbard model's simplicity, numerical attempts to verify the
properties of a Kosterlitz-Thouless (KT) transition are inconsistent
\cite{hubbkt}.
The main reason seems to be the fairly small lattice size restriction
imposed by conventional fermionic Monte Carlo methods.
Since the meron-cluster algorithm scales very efficiently with lattice
volume, we are able to go to large system sizes needed to accurately
test finite size scaling formulas.
Another advantage comes from performing the simulation directly in the
fermion occupation basis, thus allowing very simple access to a wide range
of observables.

\section{The Model}

We study a model of fermions with spin on a lattice in two spatial dimensions.
The Hamilton operator can be written as
\begin{equation}
H = \sum_{<x,y>} \;\; h_{xy} + \sum_x \;\; h_x \;,
\end{equation}
with a nearest neighbor interaction
\begin{eqnarray}
 h_{xy}&=&\sum_s \; (c_{x,s}^\dagger c_{y,s} 
  + c_{y,s}^\dagger c_{x,s} )
\nonumber\\
 &\times&\left[ (n_{xy}-1)(n_{xy}-3) + \Delta (n_{xy}-2) \right]
\nonumber \\
 &+&2 (1+\Delta) \left[\vec S_x \cdot \vec S_y +
 \vec J_x \cdot \vec J_y \right]
 - 4\Delta J_x^3 J_y^3
\nonumber \\
 &-&4\left[(n_{x,\uparrow}-\frac12)(n_{x,\downarrow}-\frac12)
  +\frac\Delta4\right]
\nonumber\\
 &\times&\left[(n_{y,\uparrow}-\frac12)(n_{y,\downarrow}-\frac12)
  +\frac\Delta4\right] \;,
\end{eqnarray}
and an on-site term
\begin{equation}
h_x = 
 -U \; (n_{x,\uparrow}-\frac12)(n_{x,\downarrow}-\frac12)
 \; - \mu \; n_x \;.
\end{equation}
We also define the number operators
\begin{eqnarray}
n_{x,s} &=& c_{x,s}^\dagger \; c_{x,s} \;\;\;\;\;
 (s\,=\,\uparrow,\downarrow)\;,
 \nonumber \\
n_x &=& n_{x,\uparrow} \;+\; n_{x,\downarrow} \;,\;\;
 n_{xy} = n_x + n_y \;,
\end{eqnarray}
along with the on-site spin
\begin{eqnarray}
\vec S_x = \frac12 \; c_{x,s}^\dagger \; \vec\sigma_{s,s'} \; c_{x,s'}
\end{eqnarray}
and pseudo-spin operators
\begin{eqnarray}
J_x^+&=&(-1)^x c_{x,\uparrow}^\dagger\;c_{x,\downarrow}^\dagger\;,\;\;
 J_x^-=(J_x^+)^\dagger \;, \nonumber\\
J_x^3&=&\frac12 \left[ n_x^\uparrow + n_x^\downarrow - 1 \right] \;.
\end{eqnarray}
The term containing $U$ serves to bind the fermions with opposite spin together
on a lattice site.
In the limit as $U\rightarrow\infty$ the model becomes bosonic and can be
mapped onto the anisotropic quantum Heisenberg model.
As $U$ decreases the model becomes increasingly fermionic, as in the
attractive Hubbard model.
However, at $U=0$ we still have a strongly interacting system due to the
additional terms present.

\begin{figure}
\begin{center}
\includegraphics*[width=0.45\textwidth]{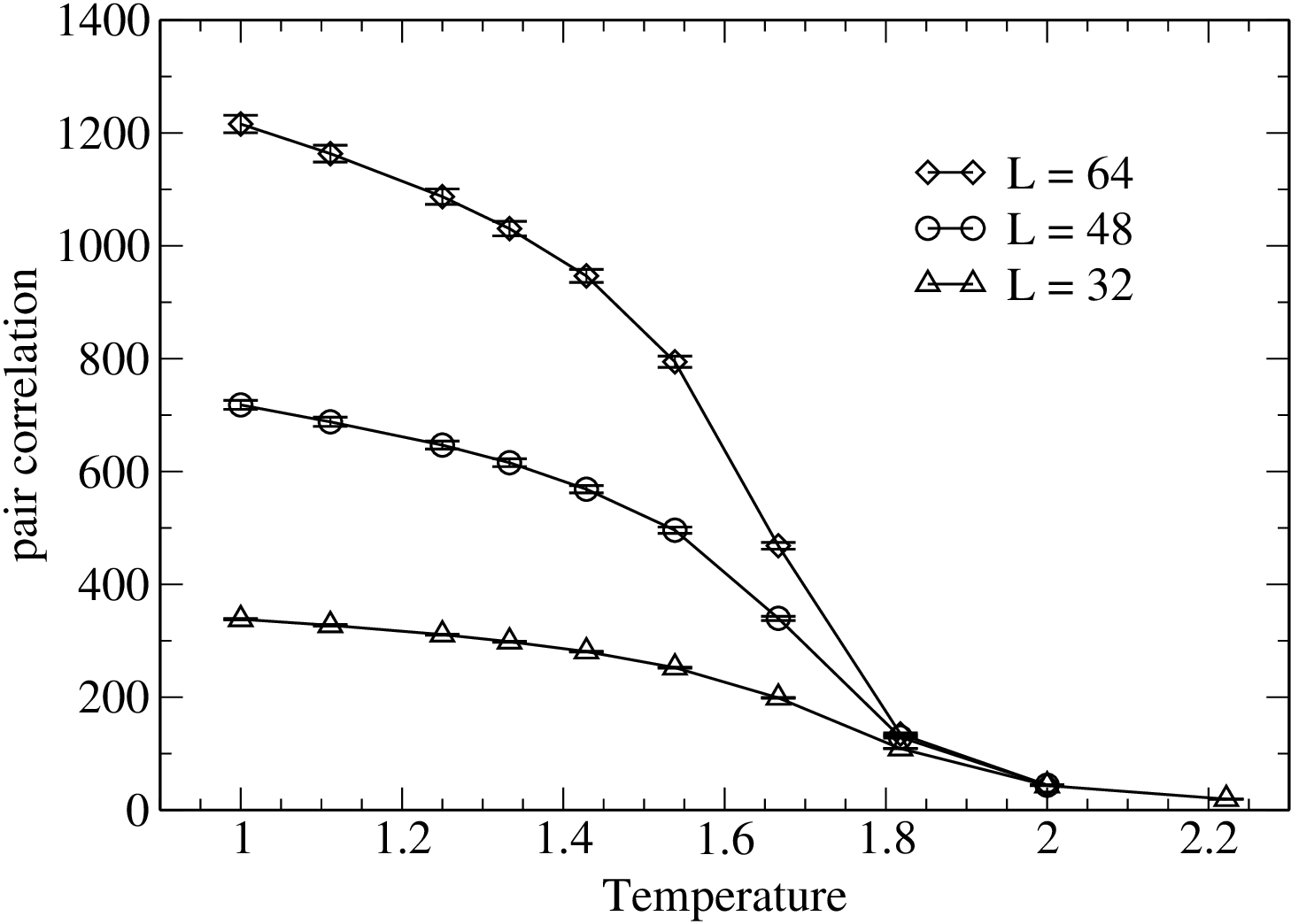}
\end{center}
\vskip-1.2cm
\caption{Pair correlation versus temperature for $U=8$ and $\Delta=1$.}
\vskip-0.4cm
\label{pcu8}
\end{figure}

\begin{figure}
\begin{center}
\includegraphics*[width=0.45\textwidth]{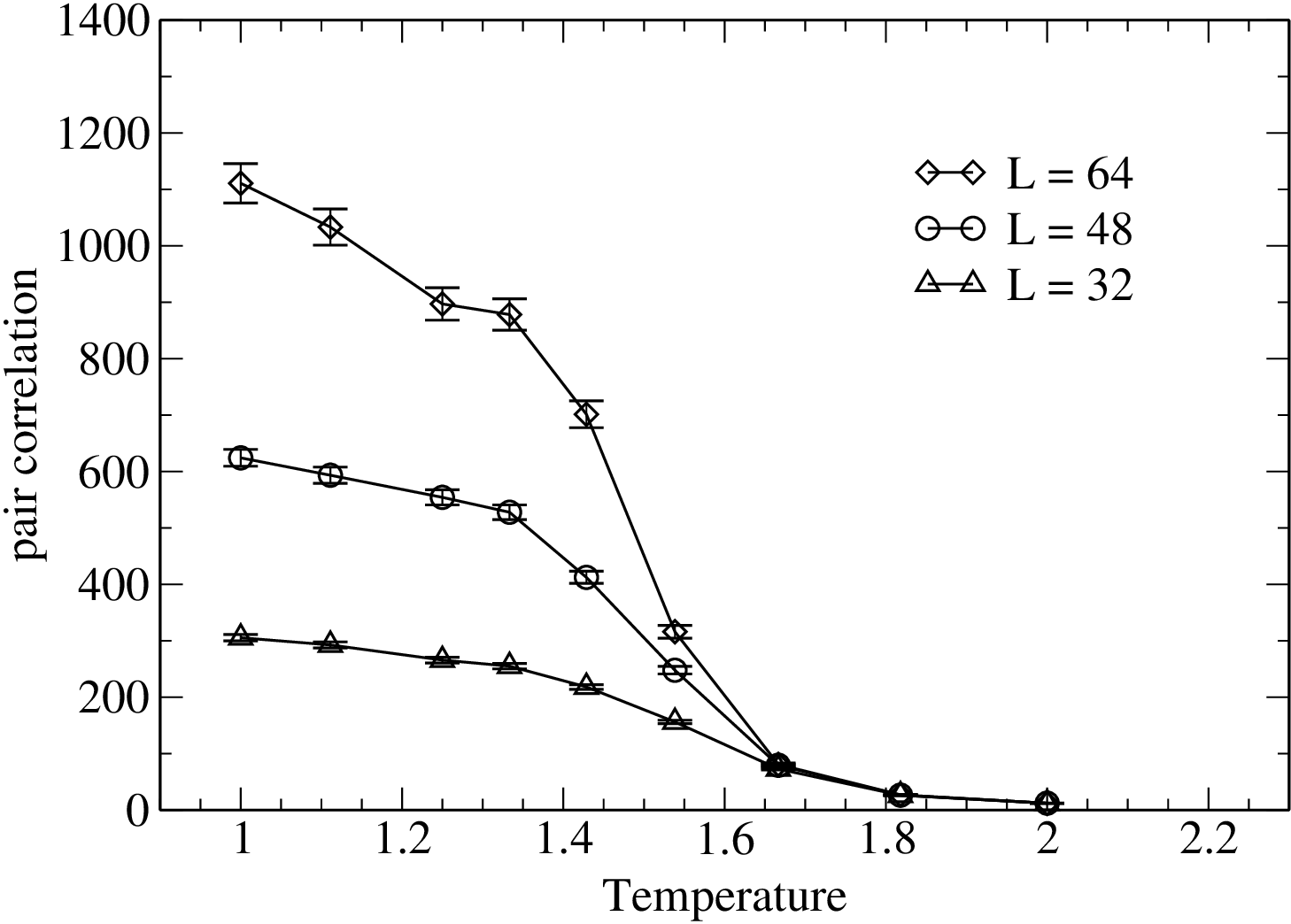}
\end{center}
\vskip-1.2cm
\caption{Pair correlation versus temperature for $U=0$ and $\Delta=1$.}
\vskip-0.4cm
\label{pcu0}
\end{figure}

In spite of these additional factors, we still expect this model to behave
similarly to the attractive Hubbard model since they both possess the same
symmetries.
The important symmetries are the $SU(2)_s$ spin group generated by the 
standard fermionic spin operator
$\vec S = \sum_x \vec S_x$,
and the $SU(2)_c$ charge symmetry generated by the pseudo-spin operator
$\vec J = \sum_x \vec J_x$.
The $SU(2)_s$ group is always a symmetry of the Hamiltonian, but
the $SU(2)_c$ group is broken down to the $U(1)_c$ particle number group
if either $\mu$ or $\Delta$ are nonzero.
The remaining $U(1)_c$ symmetry can then undergo a Kosterlitz-Thouless
transition in two dimensions.

Here we will only consider the case where the $SU(2)_c$ is broken by
$\Delta\ne0$ at half filling ($\mu=0$) as a test of the model and our
simulations.
We will study the more physically relevant case of $\mu\ne0$
elsewhere \cite{hubbmu}.

\section{Simulation Method}

The meron-cluster algorithm uses clusters in the standard path integral
formulation of a quantum model to solve the sign problem arising from
fermion permutations.
It has been presented in detail for a variety of models
\cite{meron1,meron2,staggm}.
Here we will only outline the extension to this model.

We chose this particular model because of its simple cluster rules.
The clusters are formed identically for the spin up and spin down fermions.
We can thus group all the clusters into pairs that are identical except for
having different spin.
Flipping both clusters to be in the same orientation then gives an identical
fermion sign for each spin sector, thus canceling each other out.
No matter how the clusters are shaped, we can always flip them to be in
a ``reference configuration'' with both spin sectors identical, which will
always have a positive contribution to the partition function.

The on-site attraction term proportional to $U$ is implemented with an extra
on-site plaquette similar to the mass term in the previously studied
staggered fermion model \cite{staggm}.
Here, however, the extra link can either do nothing, or it can freeze two
clusters of different spin together in the same orientation.
If frozen together the clusters can still be flipped, but both spin sectors
must be flipped together.
Note that this frozen state is consistent with the reference configuration
which allows us to include it in the meron-cluster algorithm.
This also limits us to studying an on-site attraction in this model since
a repulsion would need to freeze the clusters in opposite orientations
which is contrary to the reference configuration.

\begin{figure}
\begin{center}
\includegraphics*[width=0.45\textwidth]{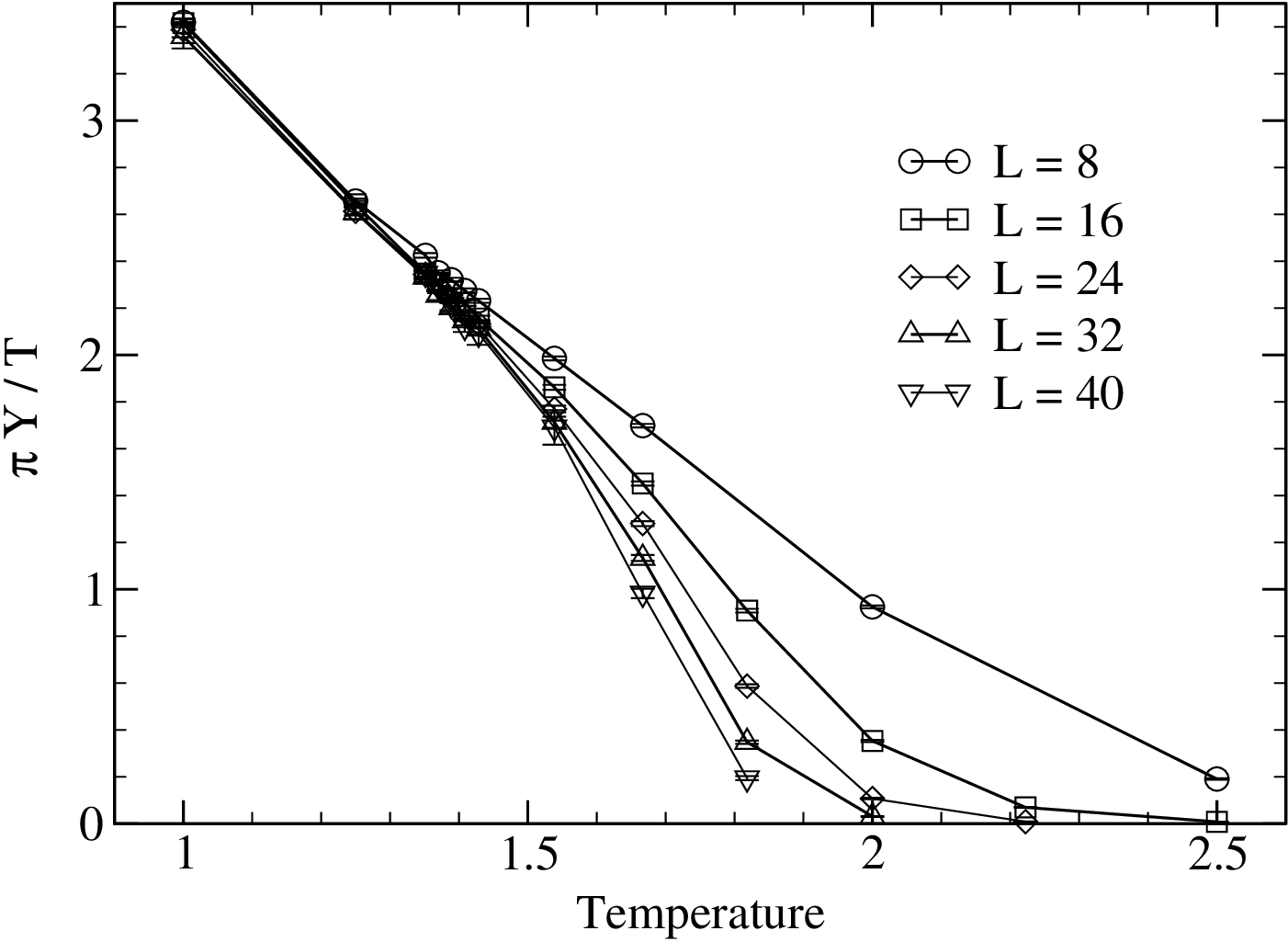}
\end{center}
\vskip-1.2cm
\caption{Winding number versus temperature for $U=8$ and $\Delta=1$.}
\vskip-0.4cm
\label{wnu8}
\end{figure}

\section{Numerical Results}

We simulated the model presented above using a variant of the implementation
used in \cite{staggm,staggi}.
We have been able to simulate systems with a spatial volume (V) of $128^2$,
but we will only present results of up to $64^2$ here.
Each simulation produced between $10^5$ and $10^6$ configurations after
performing at least $10^4$ thermalization sweeps.
In all cases we measured the autocorrelation time to be less than two
configurations.

The transition to a state with long range order can be seen in the
S-wave pair correlation
\begin{equation}
P_L \; = \;
\frac2{Z \beta V} \; \int_0^\beta \;
 {\rm Tr} \left[ \; {\rm e}^{-(\beta-t) H}\; p^- \;
 {\rm e}^{-t H}\; p^+ \;\right]
\end{equation}
with
\begin{equation}
p^+ =\;\sum_x c_{x,\uparrow}^\dagger\;c_{x,\downarrow}^\dagger \;\;, \;\;\;\;
p^- = (p^+)^\dagger \;.
\end{equation}
In the thermodynamic limit ($L\rightarrow\infty$), the pair correlation
should diverge near a KT transition as
$P_{\infty} \; \propto \; \xi_{\infty}^{\,7/4}$
with
\begin{equation}
\xi_{\infty} \; \propto \; \left\{ 
\begin{array}{ll}
  {\rm e}^{\,a\,(T-T_c)^{-1/2}} & {\rm for} \; T>T_c \\
  \infty                           & {\rm for} \; T<T_c \\
\end{array}
\right. \;.
\end{equation}

\begin{figure}
\begin{center}
\includegraphics*[width=0.45\textwidth]{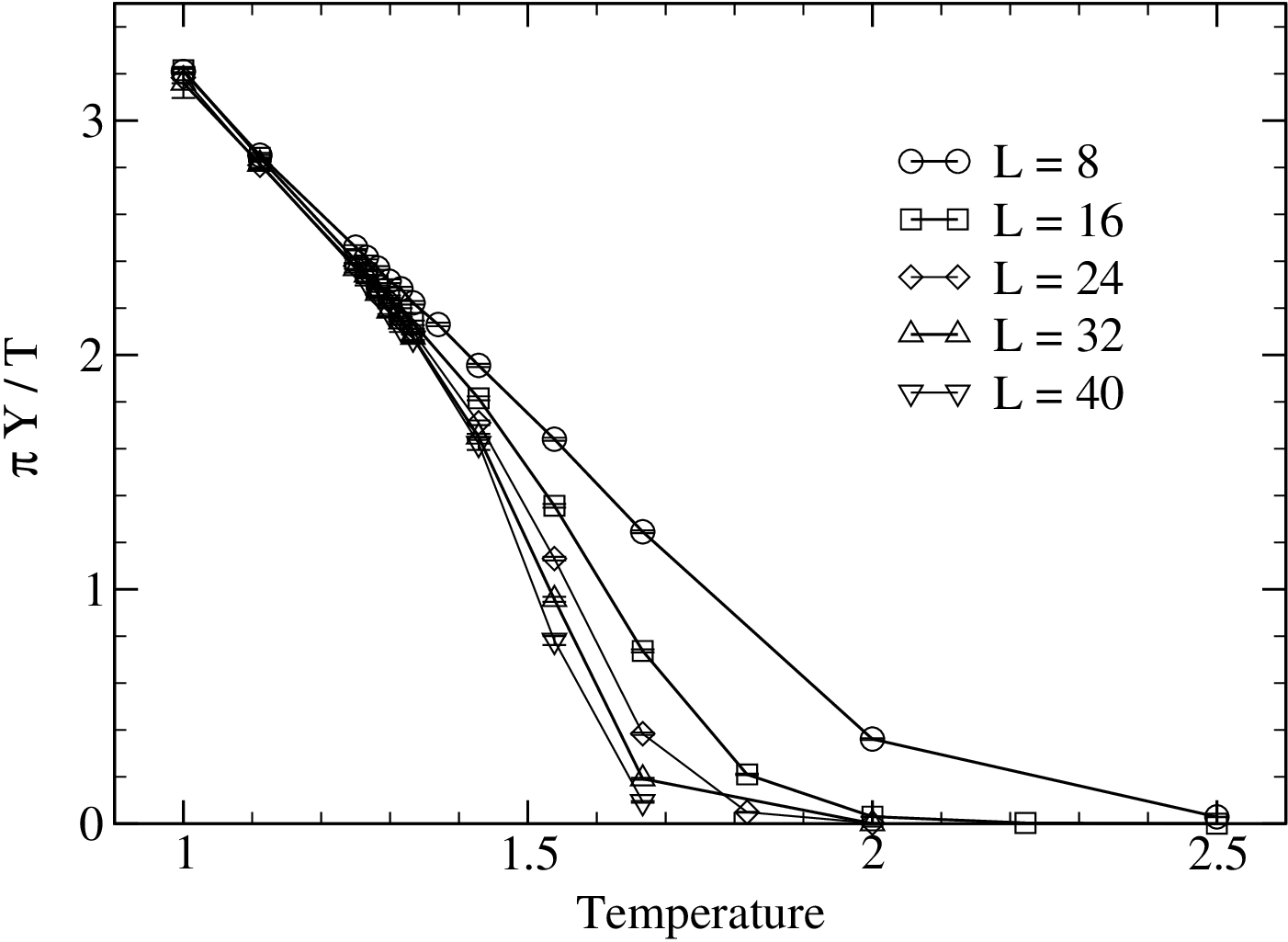}
\end{center}
\vskip-1.2cm
\caption{Winding number versus temperature for $U=0$ and $\Delta=1$.}
\vskip-0.4cm
\label{wnu0}
\end{figure}

Figures \ref{pcu8} and \ref{pcu0} show the pair correlation for
$U=8$ and $U=0$ respectively.
In both cases we can clearly see evidence of long range correlations forming.
Here we do not see much of a difference between the nearly bosonic ($U=8$) and
fermionic ($U=0$) model, only a slight shift in the critical temperature.

One could attempt to extract $T_c$ from fitting the pair correlation for the
largest volume available to the above form for $P_\infty$.
This generally turns out to be difficult due to the large volumes necessary
to approximate $P_\infty$ near $T_c$.

An easier way of measuring $T_c$ comes from the helicity modulus
\cite{helmod,Hara98} which can be defined in terms of the winding number as
\begin{equation}
\Upsilon = \frac{T}{2} \left< \; W_x^2 + W_y^2 \; \right>
\end{equation}
where $W_x$ ($W_y$) is the total number of particles winding around the
boundary in the x (y) direction.
This is very convenient to work with since we know the finite size scaling form
to be
\begin{equation}
\frac{\pi}{T}\Upsilon = A(T) \left[ 2 + \frac{1}{\mathrm{log}(L/L_0(T))}\right]
\label{fsshm}
\end{equation}
with $A(T_c)=1$.
In figures \ref{wnu8} and \ref{wnu0} we show the quantity $\pi\Upsilon/T$
for the two different values of $U$.
Here we can clearly see the tendency of a universal jump between 0 and 2
at $T_c$.
Near $T_c$ we can fit the helicity modulus for several volumes to (\ref{fsshm})
to get the two parameters $A(T)$ and $L_0(T)$.
We see that the coefficient $A(T)$ moves approximately linearly through 1 as
$T$ goes through $T_c$.
By fitting a straight line to $A(T)$ we can determine $T_c$.
For $U=8$ we estimate that $T_c=1.39(2)$ while for $U=0$ we get $T_c=1.28(2)$.

For $\Delta=1$ with $U\rightarrow\infty$ our model maps onto the quantum XY
model with an extra factor of 1/4 multiplying the temperature.
Scaling the measured result for the XY model \cite{Hara98}, we should get
$T_c=1.3708(2)$ for large enough $U$.
Our result for $U=8$ is consistent with this which provides a good check of
our simulations.
We also see that $U=8$ is indeed close to the bosonic limit.

\begin{figure}
\begin{center}
\includegraphics*[width=0.45\textwidth]{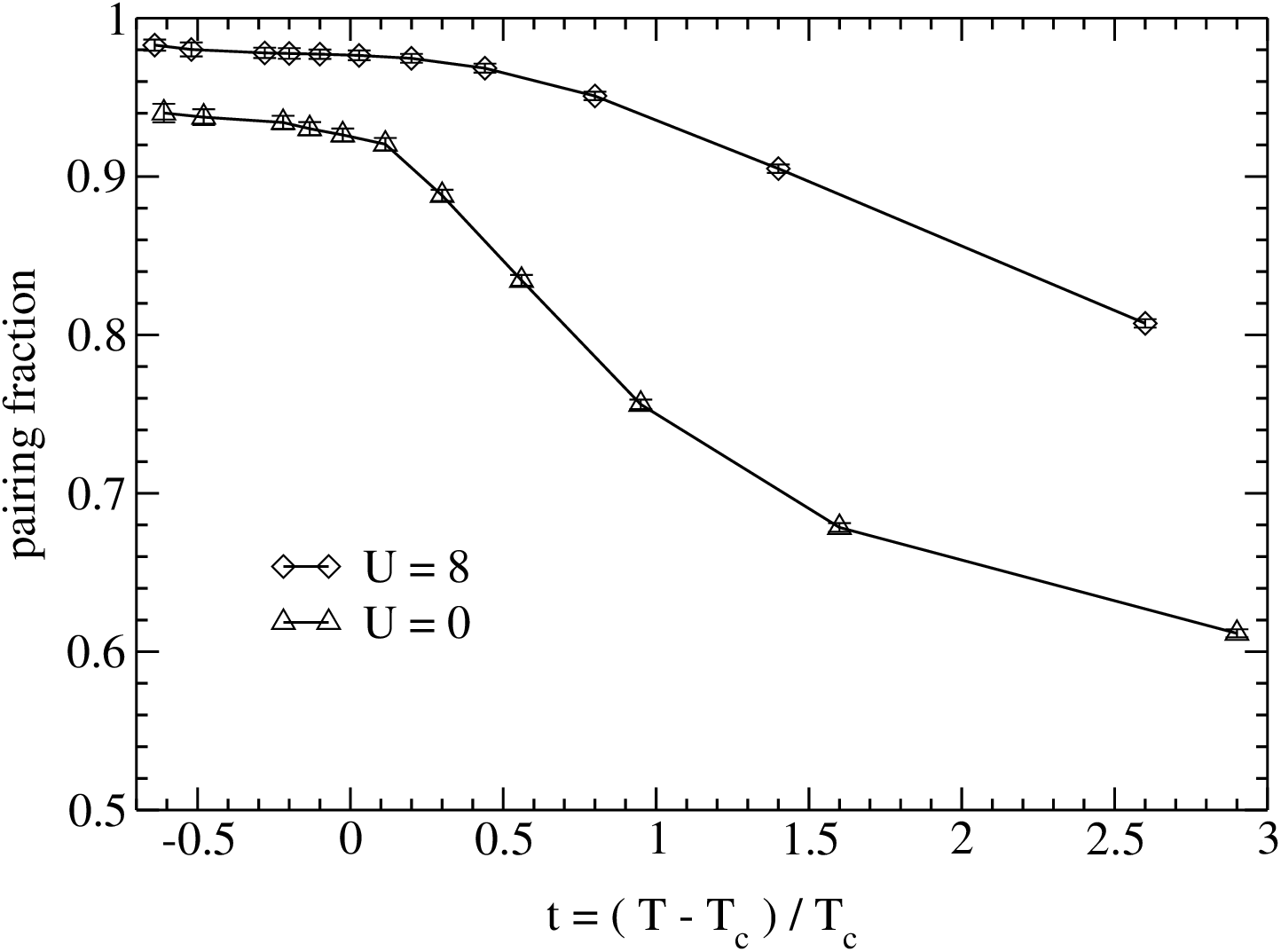}
\end{center}
\vskip-1.2cm
\caption{Pairing fraction versus reduced temperature for ($U=8$)
and fermionic ($U=0$) limits.}
\vskip-0.4cm
\label{pf32}
\end{figure}

We can better understand how the model changes as we adjust $U$
by measuring the pairing fraction which we define as
\begin{equation}
f = 2 \left< n_\uparrow n_\downarrow \right> /
 \left< n_\uparrow + n_\downarrow \right> \;.
\end{equation}
In figure \ref{pf32} we show the pairing fraction for both $U=8$ and $U=0$
versus the reduced temperature $t=(T-T_c)/T_c$ using $T_c$ as determined above.
Here we can distinctly see the effects of the on-site attraction.
For $U=8$ the pairing fraction gradually increases as $T_c$ is approached
from above and comes within a few percent of its value at $T_c$ well before
the KT transition.
For $U=0$ we see a more dramatic rise as $T_c$ is approached from above.
Here the pairing fraction levels off much closer to $T_c$ than for $U=8$,
but still just before the transition.

\section{Conclusions}

We have efficiently simulated a fermionic model with the same symmetries as
the attractive Hubbard model.
In this model we observed a KT transition at a finite
temperature for $\Delta=1$ in the bosonic (U=8) and fermionic (U=0) limits.
By measuring the helicity modulus, we were able to easily obtain
the critical temperature.

In the future we will include a chemical potential in our simulations.
This will allow us to move away from half filling.
We also plan to study this model in 3 dimensions to observe the physics of
massless Goldstone bosons.

\vspace{0.2cm}
\noindent
{\bf Acknowledgements}
\vspace{0.1cm}

We would like to thank U.-J. Wiese for motivating discussions.
A majority of the simulations were performed on computers generously donated
by Intel Corporation.

\end{document}